\documentclass[aps,prd,amssymb,amsmath,nofootinbib,twocolumn,epsf,floatfix]
              {revtex4}
\usepackage[usenames]{color} 
\usepackage[normalem]{ulem}
\usepackage{enumitem}
\usepackage{graphicx}
\usepackage{xcolor}
\usepackage{subfigure}
\usepackage{amssymb}
\usepackage{amsmath}
\usepackage{mathrsfs} 
\usepackage{epstopdf}
\usepackage{float}
\usepackage{booktabs} 
\usepackage{microtype}
\usepackage[colorlinks,breaklinks]{hyperref}
\newcommand{\be}{\begin{equation}} 
\newcommand{\ee}{\end{equation}}
\newcommand{\bea}{\setlength\arraycolsep{2pt} \begin{eqnarray}}
\newcommand{\eea}{\end{eqnarray}}

\renewcommand{\raggedright}{\leftskip=0pt \rightskip=0pt plus 0cm}

\begin{document}

\title{Nonlinear Gravitational Wave Interactions in an Expanding Universe}

\author{Fan Zhang\textsuperscript{1,2}}
\author{Lee Lindblom\textsuperscript{3}}
\author{Zhi-Peng Peng\textsuperscript{2,4} }

\affiliation{\textsuperscript{1}Institute for Frontiers in Astronomy and
  Astrophysics, Beijing Normal University, Beijing 102206, China} 
\affiliation{\textsuperscript{2}Gravitational Wave and Cosmology Laboratory,
  Department of Astronomy, Beijing Normal University, Beijing 100875, China}
\affiliation{\textsuperscript{3}Center for Dark Cosmology and Gravitation,
  Department of Physics, University of California at San Diego,
  La Jolla, CA 92093, USA}
\affiliation{\textsuperscript{4}School of Physics and Advanced Energy,
             Henan University of Technology, Henan 450001, China}
\date{\today}

\begin{abstract}
  This study explores the nonlinear interactions between gravitational
  waves in an expanding universe.  Random ensembles of gravitational
  waves are evolved numerically in cosmological models whose spatial
  volumes increase by more than an order of magnitude during these
  simulations.  These evolutions show the spectra of the gravitational
  waves evolving as energy cascades away from the wavelengths at the
  peak of the initial spectra toward both longer and shorter
  wavelengths.  Evolutions performed at different gravitational wave
  amplitudes are shown to produce cascades that scale with amplitude
  in the way predicted by the four-wave scattering models of
  gravitational-wave turbulence.
\end{abstract}

\maketitle

\section{\label{sec1}Introduction}

Gravitational waves are fluctuations in the geometry of spacetime that
propagate at the speed of light according to Einstein's general
relativity theory.  It has long been assumed that the non-linearity of
Einstein's equation could in suitable circumstances lead to
interactions between these waves which lead to turbulent cascades that
transform their spectra. These non-linear interactions were studied
first in the context of anti-de Sitter spacetime.  Using both
numerical and analytical methods the studies reported in
Refs.~\cite{Bizon2011, Dias2012} found that a turbulent cascade of
gravitational waves from lower to higher frequencies could drive an
instability in anti-de Sitter spacetime. Those studies speculate that
this instability could result in the production of black holes.  The
studies reported in Refs.~\cite{Green2014, Ma2026} have also found,
using analytical and numerical methods, that inverse turbulent
cascades of gravitational waves from higher to lower frequencies could
occur in the gravitational waves emitted by perturbed black holes.

Several studies have considered the question whether gravitational
wave turbulence could be playing a role in the evolution of our
universe, e.g., Refs.~\cite{Rica2016, Galtier2017, Clough2018,
  Galtier2020, Gay2024, Florio2026}.  The most fundamental of these
studies, Refs.~\cite{Rica2016, Galtier2017, Gay2024}, used analytical
methods to show that gravitational waves in general relativity theory
can exhibit turbulent behavior leading to energy transfers from lower
to higher frequency waves, and inverse transfers from higher to lower
frequencies under suitable circumstances.  The issue of whether such
turbulent cascades play a role in our universe is still an open
question. The analysis in Ref.~\cite{Clough2018} showed that the
expansion time in our universe is shorter than the estimated growth
time needed to establish a turbulent cascade, suggesting that
gravitational wave turbulence does not play an important role in our
universe.  A different analysis in Ref.~\cite{Galtier2020} suggests
that an inverse cascade from higher to lower frequency gravitational
waves could drive an inflation epoch in our early universe.  Those
gravitational-wave turbulence studies have all been analytical.  A
recent study however, Ref.~\cite{Florio2026}, has reported the
development and testing of numerical methods that may soon allow some
of these questions to be explored in ways that are not amenable to
analytical methods.

This paper presents the results of fully non-linear three-dimensional
numerical evolutions of gravitational waves in expanding cosmological
models.  The initial data for these evolutions were constructed on
spatial hypersurfaces with the topology of the three-torus.  The
initial data used in this study includes an ensemble of gravitational
waves (constructed using the appropriate transverse-traceless tensor
harmonics on the three-torus) having random phases and a spectrum that
peaks for wavelengths half the size of the universe.  Initial data
were constructed for ensembles of gravitational waves with three
different dimensionless amplitude scales, $\mathcal{A}_s=\{0.001,
0.01, 0.1\}$, so that the effects of non-linearity could be assessed.
The methods used to construct these initial data are described in
detail in Sec.~\ref{InitialData}.  The properties of the
transverse-traceless tensor harmonics on the three-torus are
summarized in Appendix~\ref{s:TTHarmonics}.

A covariant symmetric-hyperbolic representation of Einstein's vacuum
equation is used in this study to evolve these initial data until the
spatial volumes of the solutions have expanded by at least an order of
magnitude.  The numerical methods used to solve these evolution
equations are described in some detail in Sec.~\ref{s:Spacetime
  Evolutions}.  This section presents visualizations of some of the
results, and demonstrates that the numerical evolutions presented here
are numerically convergent.  Details of the covariant
symmetric-hyperbolic representation of Einstein's equation are
summarized in Appendices~\ref{s:DynamicalFieldTransformations} and
\ref{s:SymmetricHyperbolicSystem}.

The evolution of the spectra of the gravitational waves in these
numerical solutions are examined in Sec.~\ref{s:Spectral Evolutions}.
These results show clear evidence for the transfer of energy from
longer wavelengths into shorter wavelength gravitational waves, and
thus evidence for the initial stages of a standard turbulent cascade.
These results also show some evidence for an inverse cascade of energy
from higher to lower frequency gravitational waves.  The implications
of these results are discussed in Sec.~\ref{s:Discussion}.  For
example, it is shown that the amplitudes of the gravitational waves
created by these turbulent cascades scale with amplitude in precisely
the way predicted by the four-wave scattering models of gravitational
wave turbulence.  The more subtle evidence produced by this study for
the existence of an inverse cascade of energy from higher to lower
frequency gravitational waves is also discussed in
Sec.~\ref{s:Discussion}

\section{Initial Data}
\label{InitialData}

Solutions to Einstein's equation are constructed for this study by
evolving suitable initial data specified on a spacelike hypersurface.
The required initial data consist of the metric, $g_{ij}$, and
extrinsic curvature, $K_{ij}$, of that initial hypersurface. (The
tensor indices $i$, $j$, $...$ range over the three coordinates on
each $t=$ constant hypersurface.)  Certain components of the Einstein
equation are constraints on these fields, generally referred to as the
Hamiltonian and momentum constraints.  For a vacuum spacetime, these
constraint equations are given by~\cite{Isenberg2014},
\begin{eqnarray}
  R+K^2-K_{ij}K^{ij}&=&0
  \label{e:HamiltonianConstraint},\\ 
  D^{j}(g_{ij}K-K_{ij})&=&0
  \label{e:MomentumConstraint}.
\end{eqnarray}
Here, $R=g^{ij}R_{ij}$ is the trace of the Ricci tensor determined by
the spatial metric $g_{ij}$, $K=g^{ij}K_{ij}$ the trace of the
extrinsic curvature, and $D_{i}$ the covariant derivative compatible
with $g_{ij}$: $D_m g_{ij}=0$.

This study is focused on the interactions among an ensemble of
gravitational waves in an expanding universe. The initial data
representing these waves are constructed for this study by solving
Eqs.~(\ref{e:HamiltonianConstraint}) and (\ref{e:MomentumConstraint})
using the constant mean curvature form of the conformal initial value
problem, see e.g.  Refs.~\cite{Isenberg1995, Isenberg2014}.  In this
approach the trace of the extrinsic curvature, $K$, is assumed to be
constant on the initial hypersurface.  In this case the spatial metric
$g_{ij}$ and extrinsic curvature $K_{ij}$ can be expressed in terms of
the conformal fields $\varphi$, $e_{ij}$ and $A_{ij}$:
\begin{eqnarray}
g_{i j} &=& \varphi^4 e_{ij},
     \label{e:gij}\\
K_{i j} &=& \tfrac{1}{3}\varphi^4 e_{i j} K+\varphi^{-2}A_{ij},
     \label{e:Kij}
\end{eqnarray}
where $e_{ij}$ is the conformal metric, $\varphi$ the conformal
factor, and $A_{ij}$ the transverse-traceless part of the extrinsic
curvature.  The momentum constraint forces the longitudinal part of
the trace-free extrinsic curvature to vanish on the compact spatial
hypersurfaces used in this study~\cite{Isenberg1995}.
 
The conformal metric, $e_{ij}$, can be chosen freely in this version
of the initial value problem.  The compact orientable cosmological
models studied here are constructed on manifolds with spatial
hypersurfaces with the topology of the three-torus, $T^3$.  Therefore
the conformal metric, $e_{ij}$, can be chosen to be the flat Euclidean
metric.  This study represents $T^3$ using periodic Cartesian
coordinates, and in these coordinates $e_{ij}$ is just the identity
matrix $e_{ij}=\delta_{ij}$.  The notation, $\partial_m$, is used for
the covariant derivative compatible with the Euclidean metric,
$\partial_m\,e_{ij}=0$, which is just the partial derivative in
Cartesian coordinates.

The transverse-traceless extrinsic curvature, $A_{ij}$, can also be
chosen freely in this version of the initial value problem.  The
tensor harmonics, $Y^{\,+}_{\!ij,\,\mathbf{k}}$ and
$Y^{\,\times}_{\!ij,\,\mathbf{k}}\,$ provide a complete basis for the
symmetric transverse-traceless tensors on $T^3$~\cite{Peng2019}.  Therefore any
symmetric transverse-traceless tensor, including $A_{ij}$, can be
written as a linear combination of these harmonics:
\begin{equation}
  A_{ij}=\Re\left\{\sum_{\vec k}\left[\,
    \mathcal{A}_{\,\mathbf{k}}^{\,+}\, Y^{\,+}_{\!ij,\,\vec k}
    +\mathcal{A}_{\,\mathbf{k}}^{\,\times}\, T^{\,\times}_{ij,\,\vec k}\,\right]\right\},
  \label{e:A_HarmonicExpansion}
\end{equation}
where $\mathcal{A}_{\,\mathbf{k}}^{\,+}$ and
$\mathcal{A}_{\,\mathbf{k}}^{\,\times}$ are complex constants that
represent the spectra of the gravitational waves that contribute to
$A_{ij}$.  Explicit expressions for, and the properties of, these
tensor harmonics are given in Appendix~\ref{s:TTHarmonics}.

\begin{figure}[!h]
  \centering
  \includegraphics[height=0.35\textwidth]{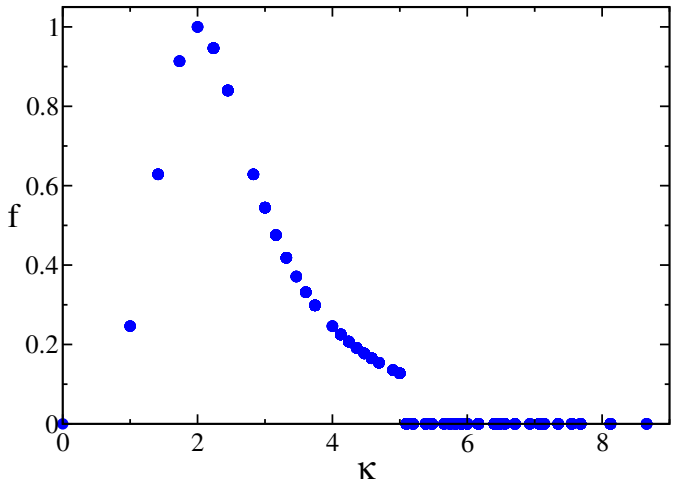}
  \caption{\label{f:InitialSpectrum} The function $f(\kappa)$
    describes the amplitudes of the gravitational-wave perturbations
    used to construct the initial data.  Each point represents the
    amplitude used for the collection of gravitational-wave modes with
    $\kappa^2=|\mathbf{k}|^2=k_x^2+k_y^2+k_z^2$, where $k_x$, $k_y$, $k_z$ are
    integers in the ranges $-5\leq k_i \leq 5$.}
\end{figure}

The extrinsic curvature, $K_{ij}$, is the Lie derivative of the
spatial metric along the timelike normals to the initial hypersurface,
and (roughly speaking) $A_{ij}$ represents the time derivative of the
propagating gravitational-wave degrees of freedom of the metric.  The
tensor harmonics, $Y^{\,+}_{\!ij,\,\mathbf{k}}$ and
$Y^{\,\times}_{\!ij,\,\mathbf{k}}\,$, multiplied by
$e^{\,i\,2\pi\,\kappa\, t}$, where
$\kappa^2=|\mathbf{k}|^2=k_x^2+k_y^2+k_z^2$, are solutions to the
linearized Einstein equation that represent small amplitude
gravitational-wave perturbations on a flat $T^3$.  The amplitudes of
the tensor harmonics $\mathcal{A}_{\,\mathbf{k}}^{\,+}$ and
$\mathcal{A}_{\,\mathbf{k}}^{\,+}$ that determine $A_{ij}$ are
therefore $2\pi\,\kappa$ times the amplitudes of these
gravitational-wave perturbations.  The spectrum of amplitudes chosen
for this study are given by,
\begin{eqnarray}
  \mathcal{A}_{\,\mathbf{k}}^{\,+} &=&
  2\pi\,\kappa\,\mathcal{A}_s\, f(\kappa)\,
  e^{\,i\,\Psi^{\,+}_{\,\mathbf{k}}}, 
  \label{e:PlusAmplitudeDef}\\
  \mathcal{A}_{\,\mathbf{k}}^{\,\times}
  &=& 2\pi\,\kappa\,\mathcal{A}_s\, f(\kappa)\,
  e^{\,i\,\Psi^{\,\times}_{\,\mathbf{k}}},
  \label{e:CrossAmplitudeDef}
\end{eqnarray}
where $f(\kappa)$ determines the spectra of the initial
gravitational-wave ensemble, $\mathcal{A}_s$ determines the overall
amplitude scale of the ensemble, and
$0\leq\Psi^{\,+}_{\,\mathbf{k}}\leq 2\pi$ and $0\leq
\Psi^{\,\times}_{\,\mathbf{k}}\leq 2\pi$ are randomly chosen phase
parameters.  The spectral function $f(\kappa)$ used in this study has
the form
\begin{equation}
  f(\kappa) = \frac{16\,\kappa^3}{64+\kappa^6},
  \label{e:fSpectrumDef}
\end{equation}
for $\kappa\leq 5$ and $f(\kappa)=0$ for $\kappa>5$.  This conformal
initial spectrum, $f(\kappa)$, which peaks at the wavelength equal to
half the size of the three-torus initial surface, is illustrated in
Fig.~\ref{f:InitialSpectrum}.

When expressed in terms of the conformal fields defined in
Eqs.~(\ref{e:gij}) and (\ref{e:Kij}), the momentum constraint,
Eq.~(\ref{e:MomentumConstraint}), is automatically satisfied for any
transverse-traceless $A_{ij}$, and therefore for any $A_{ij}$ given by
Eq.~(\ref{e:A_HarmonicExpansion}). The Hamiltonian constraint,
Eq.~(\ref{e:HamiltonianConstraint}), becomes an elliptic partial
differential equation that determines the conformal factor $\varphi$:
\begin{equation}
  \partial^m\partial_m\,\varphi =
  \tfrac{1}{12} K^2 \varphi^{\,5} - \tfrac{1}{8} A_{ij}A^{ij} \varphi^{-7}.
  \label{e:varphiEqn}
\end{equation}
The integral of the left side of Eq.~(\ref{e:varphiEqn}) vanishes on
the $T^3$ initial hypersurface used in this study, so the integral of
the right side must also vanish, which implies
\begin{equation}
  K^2 \int \varphi^{\,5}\, d^{\,3}x
  = \tfrac{3}{2}\int A_{ij}\,A^{ij}\,\varphi^{-7}\,d^{\,3}x.
  \label{e:TraceKConstraint}
\end{equation}
This condition determines $K$ as a function of the conformal factor
and must be imposed together with Eq.~(\ref{e:varphiEqn}) to determine
$\varphi$. 

Solutions to these equations are found numerically for this study
using the pseudo-spectral methods implemented in the \verb!SpEC!
  numerical relativity code (developed initially by the Caltech and
  Cornell numerical relativity groups)~\cite{Pfeiffer2003,
    Lindblom2022a}. These numerical solutions are carried out by
  \verb!SpEC! using the ksp linear and the snes non-linear solvers
  from the \verb!PETSC!  software library~\cite{petsc-user-ref}.
  These methods solve the non-linear Eq.~(\ref{e:varphiEqn})
  iteratively by minimizing a sequence of discrete versions of the
  residual $\mathcal{E}_{(n)}$ defined by
\begin{equation}
  \mathcal{E}_{(n)}=\partial^m\partial_m\varphi_{(n)}
  -\tfrac{1}{12}K^2_{(n-1)}\varphi^5_{(n)}+\tfrac{1}{8}A_{ij}A^{ij}\varphi^{-7}_{(n)},
  \label{e:CMCResidualDef}
\end{equation}
where $\varphi_{(n)}$ is the function that minimizes
$\mathcal{E}_{(n)}$, and $K_{(n-1)}$ is determined by
Eq.~(\ref{e:TraceKConstraint}) using $\varphi_{(n-1)}$ from the
previous iteration.  The calculations were initiated in this study by
setting $\varphi_{(0)}=1$, then computing $K_{(0)}$ from
Eq.~(\ref{e:TraceKConstraint}), and iterating the process until
convergence of the minimized $\mathcal{E}_{(n)}$ is achieved.

\begin{figure}[!t] 
  \centering  
  \subfigure{
    \includegraphics[height=0.4\textwidth]{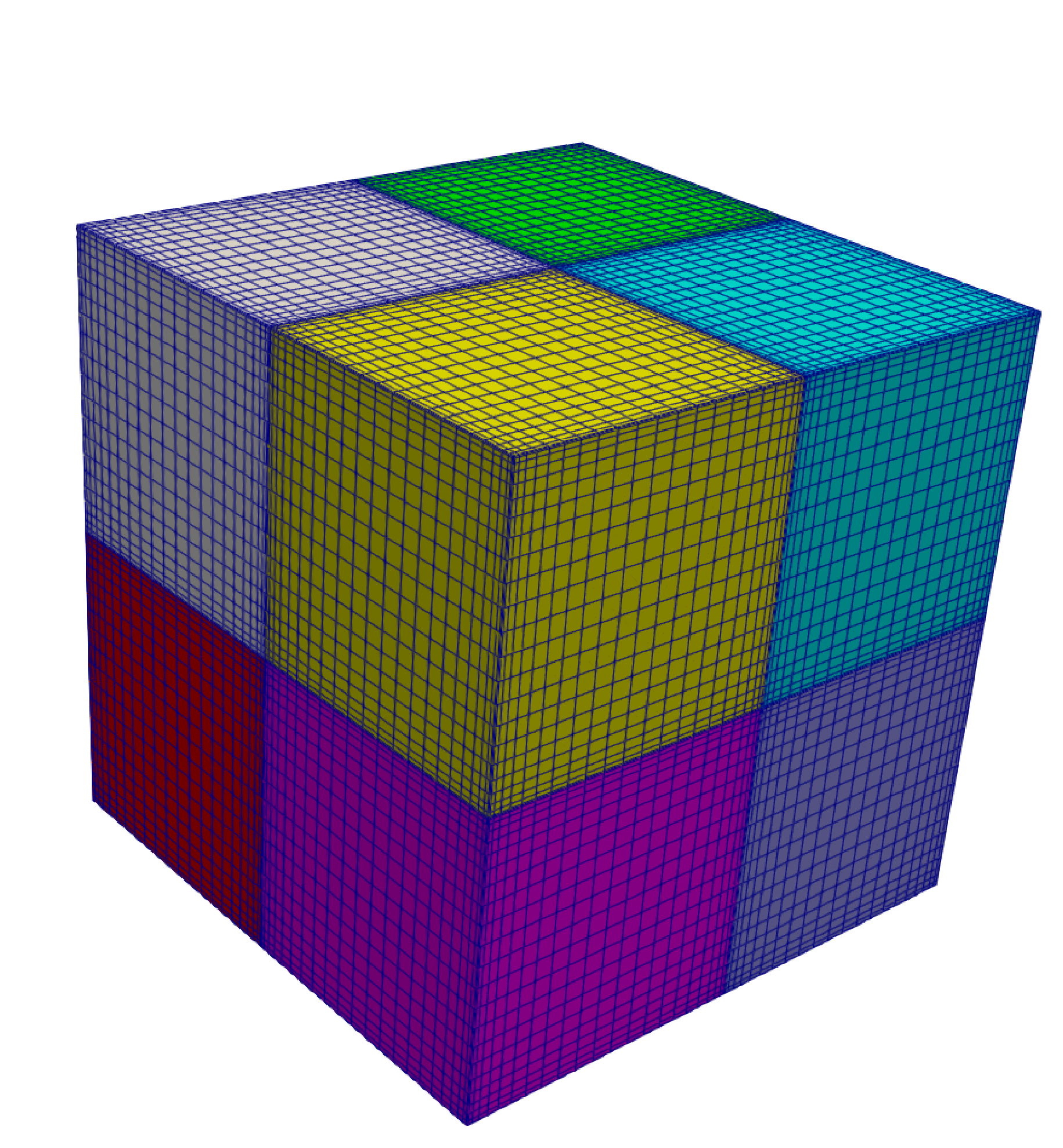}
  }
  \caption{\label{f:T3GridStructure} This figure illustrates the
    computational domain used in this study.  The tree-torus topology is
    implemented by identifying points on the opposite faces of the
    cubic domain, which is subdivided into eight smaller cubic
    subdomains.  The pseudo-spectral grid is illustrated here for the
    lowest resolution used in this study, with $N=24$
    grid points in each direction in each subdomain.}
\end{figure}

The numerical computations for this study were carried out on the
cubic computational domain illustrated in
Fig.~\ref{f:T3GridStructure}.  The three-torus topology of this domain
is implemented by imposing boundary conditions that identify points on
the opposite faces of the cubic domain.  The computations on this
domain are subdivided into the eight cubic subdomains shown in
Fig.~\ref{f:T3GridStructure} for computational efficiency and to achieve
higher resolutions.  Pseudo-spectral grids are constructed on each
subdomain with $N$ collocation points in each spatial dimension in
each subdomain.  Computations for this study were carried out on grids
with resolutions $N=\{24,32,40,48\}$.

\begin{figure}[!h] 
  \centering  
  \subfigure{
    \includegraphics[height=0.4\textwidth]{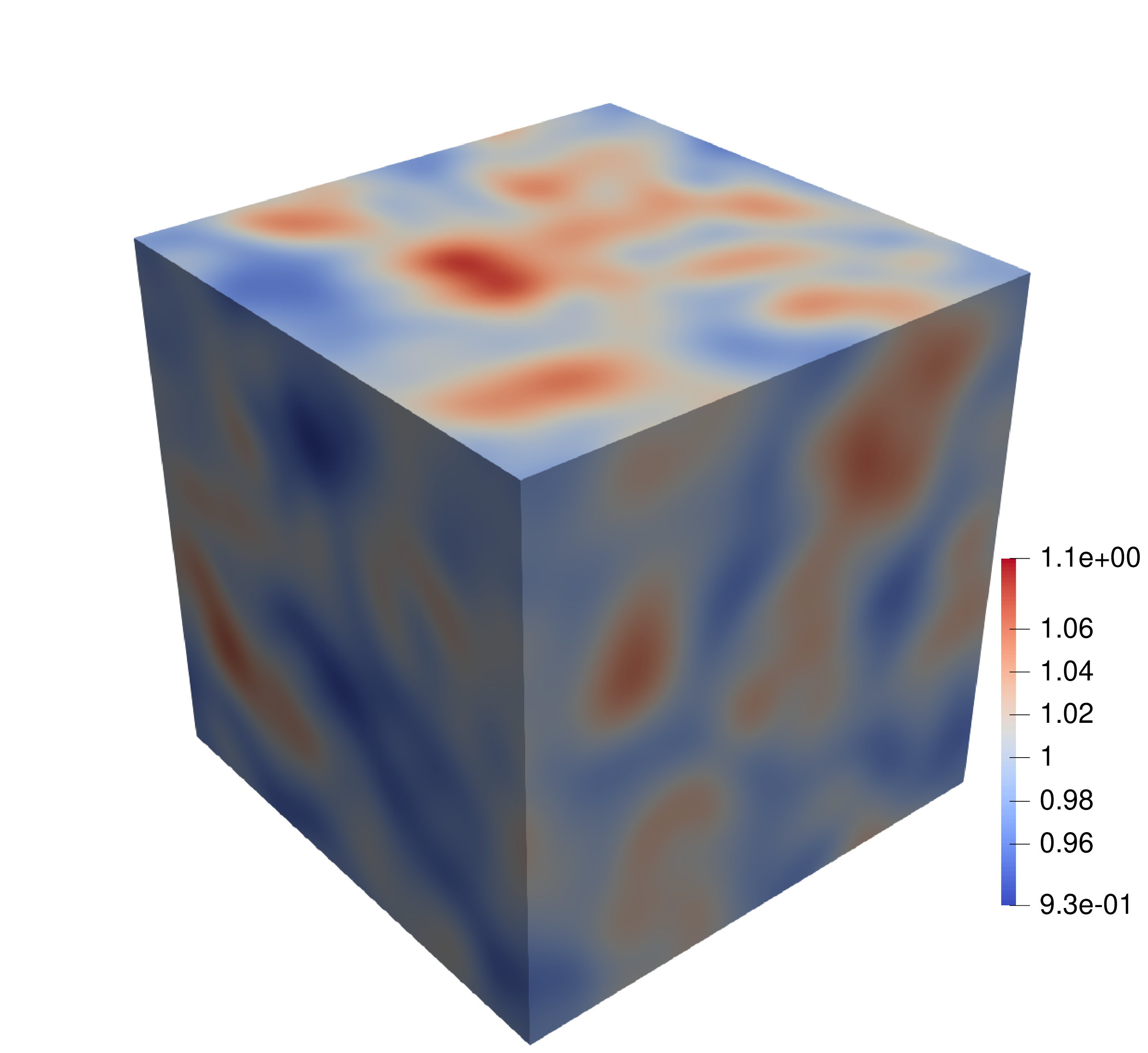}
  }
  \caption{\label{f:Amp0.1phi} This figure illustrates the solution of
    the initial value problem, $\varphi$, for the gravitational-wave
    amplitude scale $\mathcal{A}_s=0.1$ case.  }
\end{figure}

The initial value Eqs.~(\ref{e:varphiEqn}) and
(\ref{e:TraceKConstraint}) were solved for this study using
gravitational-wave data having three different amplitude scales:
$\mathcal{A}_s=\{0.001,0.01,0.1\}$. Figure~\ref{f:Amp0.1phi}
illustrates $\varphi$ for the solution with gravitational-wave
amplitude $\mathcal{A}_s=0.1$ and spatial resolution $N=48$.
Solutions for the other gravitational-wave amplitudes have similar
structures, but smaller variations away from unity.

The accuracy of the solutions to the initial value equations computed
for this study were measured by evaluating a norm of the Hamiltonian
constraint $\langle\mathcal{H}\rangle$, defined by
\begin{equation}
  \langle\mathcal{H}\rangle^2 = \tfrac{1}{\mathcal{V}}
  \int \sqrt{\det g} \left(R + K^2 - K_{ij}K^{ij}\right)^2 d^{\,3} x,
\label{e:HamiltonianNorm}  
\end{equation}
where the spatial volume $\mathcal{V}$ of the spacelike hypersurface is
determined by the integral
\begin{equation}
  \mathcal{V} = \int \sqrt{\det g}\, d^{\,3}x.
  \label{e:VolumeDef}
\end{equation}
On the initial hypersurface the expression for the volume,
$\mathcal{V}(0)$, reduces in Cartesian coordinates to the expression
\begin{equation}
\mathcal{V}(0) = \int \varphi^6 \,d^{\,3}x,
  \label{e:Volume0}
\end{equation}
which depends (weakly) on the gravitational-wave amplitude scale
$\mathcal{A}_s$.  The solutions to the initial value equations with
different gravitational-wave scales also result in different values
for the trace of the extrinsic curvature, $K$.  The resulting
extrinsic curvatures are roughly proportional to the wave amplitude
scale $\mathcal{A}_s$, so the expansion rates of the spacetimes for
the different cases differ by comparable amounts.

The resulting values of the Hamiltonian constraint norms,
$\langle\mathcal{H}\rangle$, the initial volumes, $\mathcal{V}(0)$,
and the trace of the extrinsic curvature $K$ are given in
Table~\ref{t:TableI} for the different gravitational-wave amplitude
scales, $\mathcal{A}_s$, included in this
study.  These
results are for the highest resolution solutions from each
gravitational-wave scale $\mathcal{A}_s$ ($N=40$ for
$\mathcal{A}_s=0.001$ and $N=48$ for $\mathcal{A}_s=0.01$ and
$\mathcal{A}_s=0.1$).
\begin{table}[!hbt]
  \caption{The initial extrinsic curvatures $K$, spatial volumes
    $\mathcal{V}$ and norms of Hamiltonian constraint,
    $\langle\mathcal{H}\rangle$ for the highest resolution solutions
    to Eqs.~(\ref{e:varphiEqn}) and (\ref{e:TraceKConstraint}) for the
    gravitational-wave amplitude scales $\mathcal{A}_s$ included in
    this study.  (This study uses the Ref.~\cite{MTW1971} sign
    convention for the extrinsic curvature $K$ in which $K<0$
    corresponds to expansion.).
    \label{t:TableI} }
  \begin{center}
  \begin{tabular}{llll} 
    \quad$\mathcal{A}_s$ & \quad\qquad $\langle\mathcal{H}\rangle$
     & \quad\qquad $\mathcal{V}(0)$ & \quad\qquad $K$  \\ 
 \hline
 \vspace{-7pt}\\
 0.001 & \qquad $3.45\times 10^{-9}$  & \qquad 1.000004 & \qquad -0.21194  \\
 0.01  & \qquad $3.64\times 10^{-8}$  & \qquad 1.000461 & \qquad -2.11445 \\
 0.1   & \qquad $2.51\times 10^{-7}$  & \qquad 1.022513 & \qquad -20.0228 \\
 \hline
  \end{tabular}
  \end{center}
\end{table}
%

\section{Spacetime Evolutions}
\label{s:Spacetime Evolutions}

The representation of the Einstein evolution equation used in this
study is the covariant first-order symmetric-hyperbolic representation
developed in Ref.~\cite{Lindblom2014}, whose salient features are
summarized in Appendix~\ref{s:SymmetricHyperbolicSystem}.  The
dynamical fields in this representation are the spacetime metric,
$\psi_{ab}$, its time derivative,
$\Pi_{ab}=-t^c\tilde\nabla_c\psi_{ab}$, and its spatial derivatives
$\Phi_{iab}=\tilde\nabla_i\psi_{ab}$, where $t^c$ is the future
directed timelike unit normal to the $t =$ constant hypersurfaces.
The indices from the first part of the alphabet, e.g. $a$, $b$, $c$, $...$,
range over the four-dimensional spacetime coordinates, while the
indices from the later parts of the alphabet, e.g. $i$, $j$, $...$, range
over the three-dimensional spatial coordinates on each $t=$ constant
hypersurface.

The covariant derivative, $\tilde\nabla_a$, used in the definitions of
the first-order dynamical fields is the one compatible with a
time-independent reference metric $\tilde\psi_{ab}$:
$\tilde\nabla_c\tilde\psi_{ab}=0$.  Another essential use of this
reference metric is the construction of the mappings that enforce the
appropriate boundary conditions at the interfaces between the cubic
regions that serve as coordinate charts on the spacelike
hypersurfaces~\cite{Lindblom2013, Lindblom2014}.  The spacetimes
included in this study have spacelike hypersurfaces with three-torus
topologies.  Therefore a natural choice for the reference metric in
this case is the flat Minkowski metric
\begin{equation}
  d\tilde s^2 = \tilde\psi_{ab}\, dx^a dx^b = -dt^2 + e_{ij}\, dx^i dx^j.
  \label{e:ReferenceMetricDef}
\end{equation}
In this study the reference co-variant derivative is written as
$\tilde\nabla_a=\partial_a$ since it reduces to partial derivatives in
Cartesian coordinates.

The spatial metrics, $g^0_{ij}$, produced by the solutions to the
Einstein constraint equations described in Sec.~\ref{InitialData} are
used to construct the initial data for the spacetime metric
$\psi^0_{ab}$:
\begin{equation}
  ds^2=\psi^0_{ab}\,dx^adx^b = - dt^2 + g^0_{ij}\,dx^idx^j,
  \label{e:psi0Def}
\end{equation}
The values of the other dynamical fields on the initial hypersurface are
given by,
\begin{eqnarray}
  \Pi^0_{ta}&=&0,
  \label{e:InitialPita}\\
  \Pi^0_{ij}&=& 2K^0_{ij},
  \label{e:InitialPiij}\\
  \Phi^0_{iab}&=&\partial_i\psi^0_{ab},
  \label{e:InitialPhi}
\end{eqnarray}
where $K^0_{ij}$ is the physical extrinsic curvature determined from
the solution to the initial value equations in Sec.~\ref{InitialData}.
A more complete discussion of the relationship between the first-order
dynamical fields $\psi_{ab}$, $\Pi_{ab}$, $\Phi_{iab}$ and the more
standard three plus one dynamical fields $g_{ij}$ and $K_{ij}$ is
given in Appendix~\ref{s:DynamicalFieldTransformations}.

The covariant symmetric-hyperbolic representation of the Einstein
evolution equation used in this study has been implemented in the
\verb!SpEC! numerical relativity code~\cite{Lindblom2014}.  This code
has been tested by performing fully nonlinear evolutions of small
amplitude gravitational wave ensembles propagating in the Einstein
static universe~\cite{Lindblom2014}.  Those tests demonstrated high
precision numerical convergence and long term numerical stability for
the evolutions lasting about 160 light-crossing times of the
three-sphere spatial slices.  This code has also been tested by
performing evolutions of expanding homogeneous and non-homogeneous
cosmological models on manifolds with spatial slices having
non-orientable topologies~\cite{Lindblom2026a}.  These non-orientable
cosmological model tests were shown to be numerically convergent for
the evolutions lasting long enough to increase their spatial volumes by
a factor of 10.

The numerical evolutions for this study were carried out using the
initial data constructed in Sec.~\ref{InitialData} with
gravitational-wave ensembles having amplitude scales
$\mathcal{A}_s=\{0.001,0.01,0.1\}$, on computational grids with
spatial resolutions $N=\{24,32,40,48\}$ in each direction in each of
the eight cubic regions.  The time integrations for the evolutions in
the present study were carried out using an eighth-order
Dormand-Prince adaptive timestep integrator~\cite{Dormand1987} that is
part of the \verb!SpEC!  numerical relativity code.  The timestep
error tolerance was set at $10^{-5-N/16}$ (where $N$ is the spatial
resolution) for the evolutions completed for this study.

\begin{figure}[!h] 
  \centering  
  \subfigure{
    \includegraphics[height=0.3\textwidth]{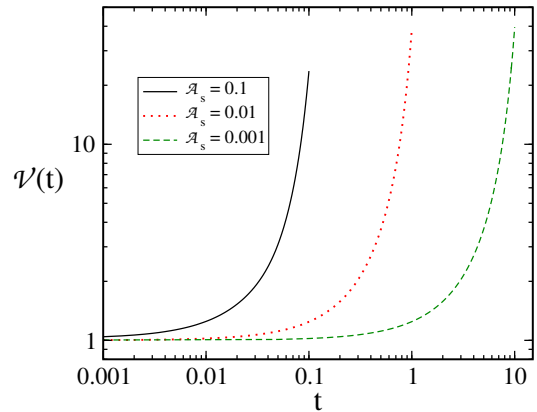}
  }
  \caption{\label{f:Volume} This figure illustrates the volumes,
    $\mathcal{V}(t)$, defined in Eq.~(\ref{e:VolumeDef}) for the
    evolutions with gravitational-wave amplitude scales
    $\mathcal{A}_s=\{0.001,0.01,0.1\}$.  }
\end{figure}

The goal of this study was to evolve initial data for ensembles of
gravitational waves until changes from the initial gravitational-wave
spectra could be observed.  To accomplish this goal, the evolutions
were carried out until the volumes of the spatial hypersurfaces of the
spacetime models grew by at least an order of magnitude.  The initial
expansion rates for these spacetime models are set by the initial
values of the trace of the extrinsic curvature $K^0$.  The values of
$K^0$ from Table~\ref{t:TableI} are roughly inversely proportional to
the gravitational-wave amplitude scales $\mathcal{A}_s$. Consequently
the coordinate time evolutions required to achieve the desired volume
growths also need to be inversely proportional to $\mathcal{A}_s$.
These considerations determined the coordinate time intervals for the
evolutions in this study: $0\leq t \leq 0.1$ for the
$\mathcal{A}_s=0.1$ case, $0\leq t \leq 1.0$ for $\mathcal{A}_s=0.01$,
and $0\leq t \leq 10.0$ for $\mathcal{A}_s=0.001$.
Figure~\ref{f:Volume} illustrates the evolution of the physical
volumes of the $t$=constant spacelike hypersurfaces $\mathcal{V}(t)$
defined in Eq.~(\ref{e:VolumeDef}) for these cases.

\begin{figure}[!h]
  \centering  
  \subfigure{
    \includegraphics[height=0.4\textwidth]{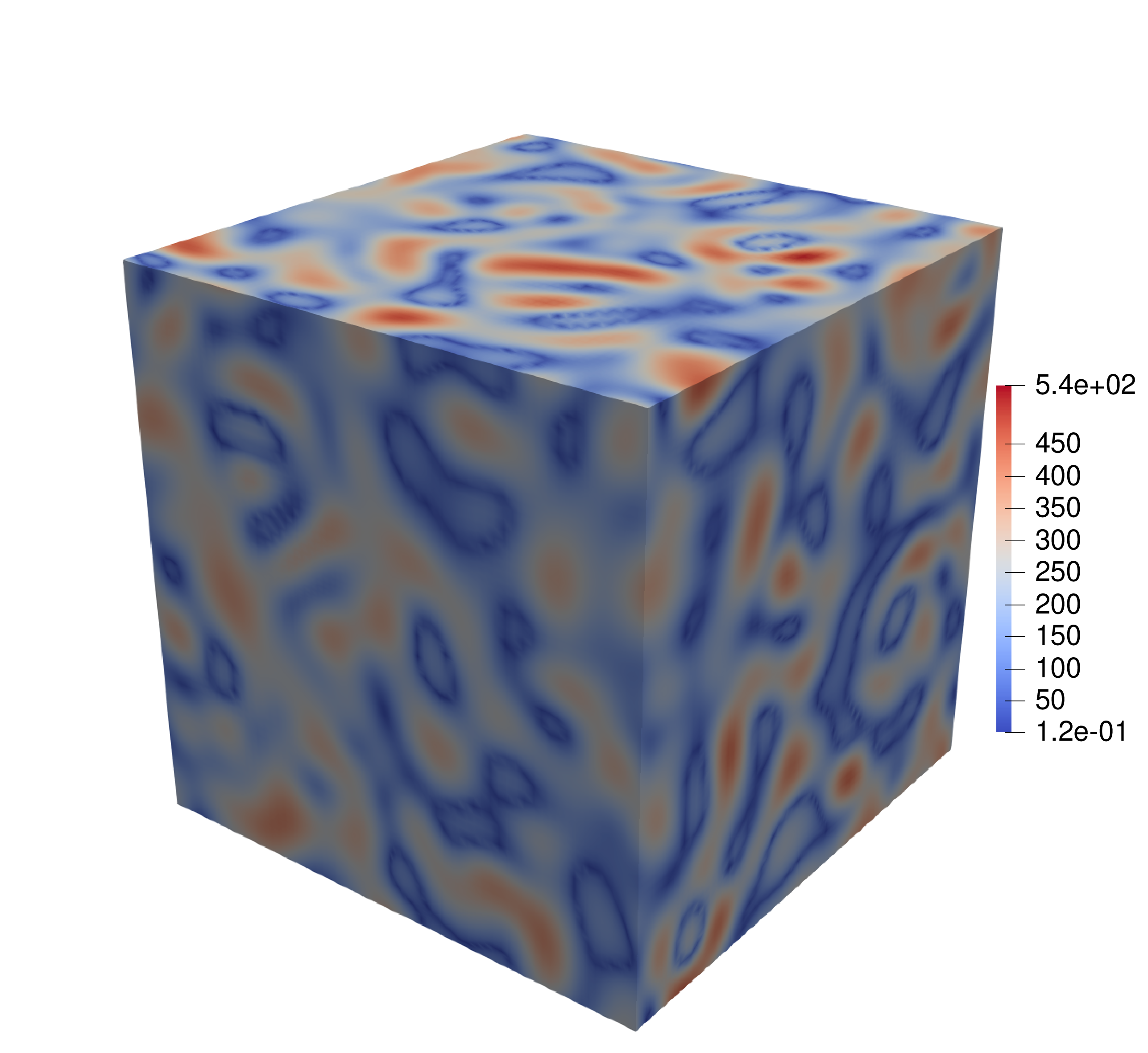}
  }
  \caption{\label{f:WeylCurvature0} This figure illustrates the Weyl
    curvature scalar $\mathcal{W}$ defined in Eq.~(\ref{e:WeylScalar})
    at time $t=0$ for the evolution of the gravitational wave ensemble
    with amplitude scale $\mathcal{A}_s=0.1$.}
\end{figure}

\begin{figure}[!h]
  \centering  
  \subfigure{
    \includegraphics[height=0.4\textwidth]{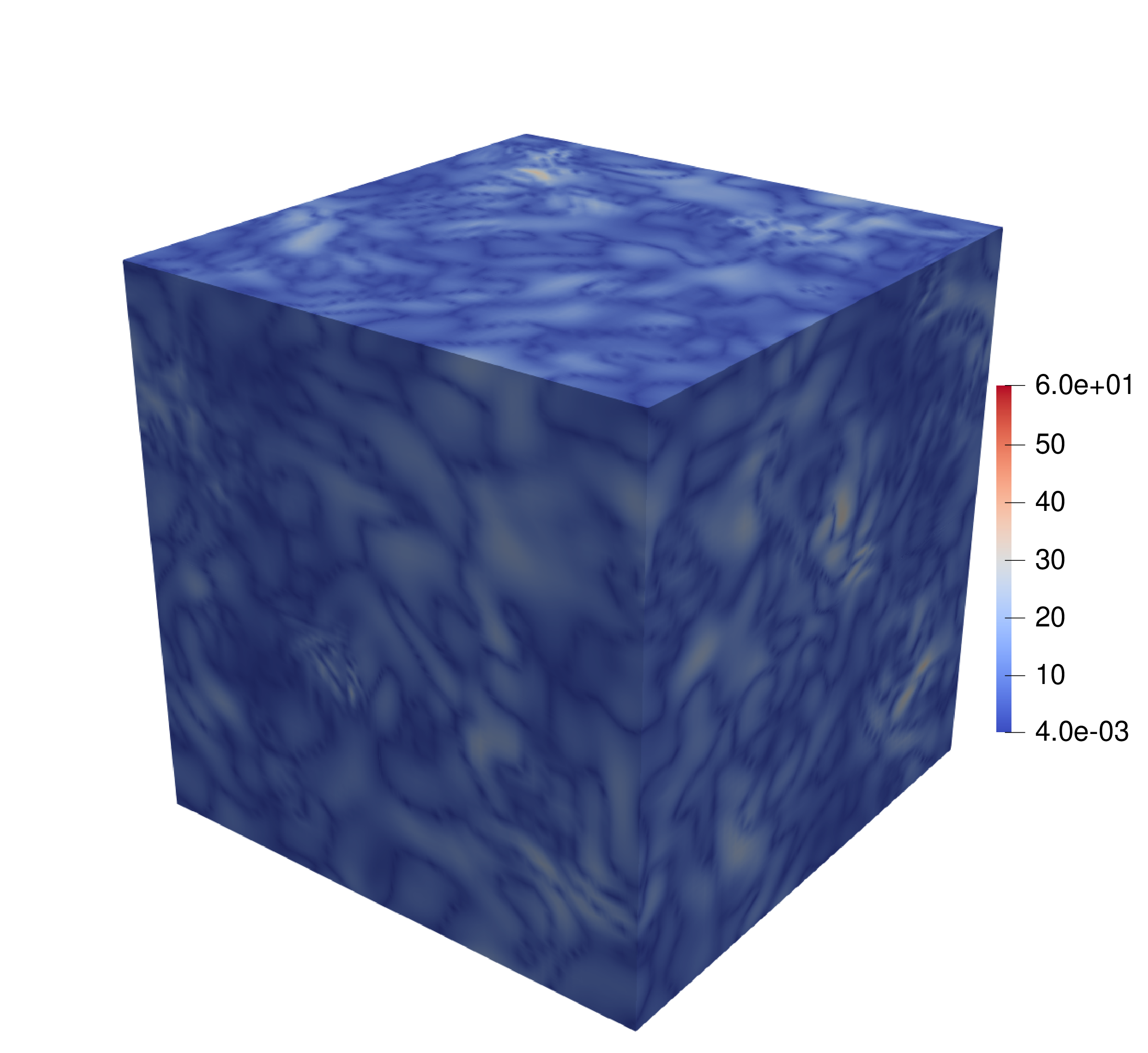}
  }
  \caption{\label{f:WeylCurvature0.1} This figure illustrates the Weyl
    curvature scalar $\mathcal{W}$ defined in Eq.~(\ref{e:WeylScalar})
    at the final time $t=0.1$ for the evolution of the gravitational
    wave ensemble with amplitude scale $\mathcal{A}_s=0.1$.}
\end{figure}

The evolution of the gravitational-wave spectra in these expanding
cosmological models is explored quantitatively in some detail in
Sec.~\ref{s:Spectral Evolutions}.  A more qualitative approach is
taken here using the Weyl tensor, which is the traditional geometrical
tool for studying the properties of gravitational waves in vacuum
spacetimes.  An impression of the evolution of the gravitational-wave
spectra in these models can be obtained by visualizing the curvature
scalar
\begin{equation}
  \mathcal{W}^2= \tfrac{1}{16}\bigl|C_{abcd}C^{abcd}\bigr|
  \label{e:WeylScalar}
  \end{equation}
where $C_{abcd}$ is the Weyl tensor.  Figures~\ref{f:WeylCurvature0}
and \ref{f:WeylCurvature0.1} illustrate $\mathcal{W}$ on the surfaces
of the computational domain for the evolution of the
gravitational-wave ensemble with amplitude scale $\mathcal{A}_s=0.1$.
Figure~\ref{f:WeylCurvature0} shows $\mathcal{W}$ at the initial
coordinate time $t=0$ and Fig.~\ref{f:WeylCurvature0.1} and at the
final time $t=0.1$.  These figures show that the overall pattern of
gravitational wave induced curvature evolves toward shorter
wavelengths and more overall spatial uniformity during the course of
this evolution.

\begin{figure}[!h] 
  \centering  
  \subfigure{
    \includegraphics[height=0.3\textwidth]{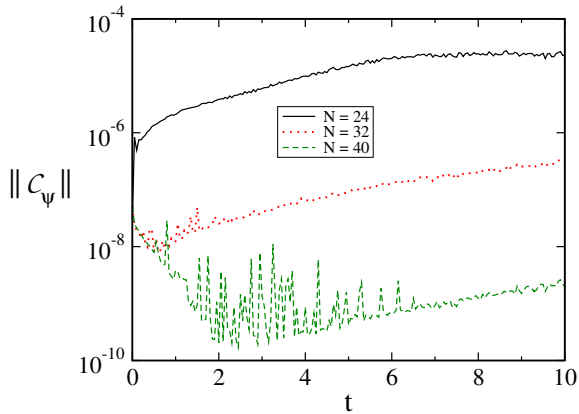}
  }
  \caption{\label{f:Amp0.001GhCeCov} This figure illustrates the
    constraint norms $||\,\mathcal{C}_\psi||$ defined in
    Eq.~(\ref{e:CpsiNormDef}) for the $\mathcal{A}_s=0.001$ evolutions.
  }
\end{figure}

\begin{figure}[!h] 
  \centering   
  \subfigure{
    \includegraphics[height=0.3\textwidth]{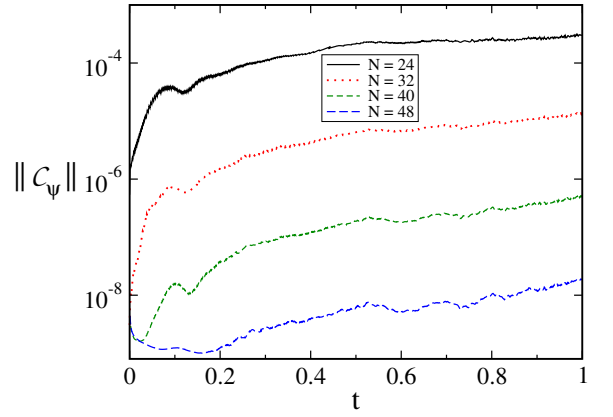}
  }
  \caption{\label{f:Amp0.01GhCeCov} This figure illustrates the
    constraint norms $||\,\mathcal{C}_\psi||$ defined in
    Eq.~(\ref{e:CpsiNormDef}) for the $\mathcal{A}_s=0.01$ evolutions.
  }
\end{figure}

\begin{figure}[!h] 
  \centering  
  \subfigure{
    \includegraphics[height=0.3\textwidth]{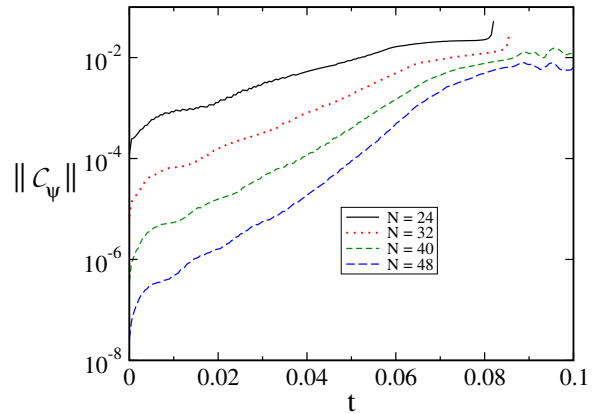}
  }
  \caption{\label{f:Amp0.1GhCeCov} This figure illustrates the constraint
    norms $||\,\mathcal{C}_\psi||$ defined in
    Eq.~(\ref{e:CpsiNormDef}) for the $\mathcal{A}_s=0.1$ evolutions.  }
\end{figure}

The first-order symmetric-hyperbolic representation of the Einstein
system has a large number of constraints, which are described in some
detail in Appendix~\ref{s:SymmetricHyperbolicSystem}.  The overall
magnitude of constraint violations is summarized by the scalar
$\mathcal{C}_\psi$, defined in Eq.~(\ref{e:CpsiDef}), that vanishes if
and only if all the constraints are satisfied.  The dimensionless norm
of this quantity $||\,\mathcal{C}_\psi\,||$, defined in
Eq.~(\ref{e:CpsiNormDef}), provides a useful measure of the fractional
constraint violation errors in the cosmological models produced by
these evolutions.  Figures~\ref{f:Amp0.001GhCeCov}, \ref{f:Amp0.01GhCeCov}, and
\ref{f:Amp0.1GhCeCov} show $||\,\mathcal{C}_\psi\,||(t)$ for the
evolutions of the initial data with gravitational-wave amplitude
scales $\mathcal{A}_s=\{0.001, 0.01,0.1\}$ respectively.  These graphs
show that the numerical methods used in this study are convergent as
the resolution $N$ of the computational grid is increased.

The evolutions of the $\mathcal{A}_s=0.001$ initial data have
significantly lower constraint violation norms
$||\,\mathcal{C}_\psi\,||(t)$ than the $\mathcal{A}_s=\{0.01,0.1\}$
evolutions.  Therefore the highest resolution, $N=48$, evolution for
the $\mathcal{A}_s=0.001$ case was not included in this study.  The
higher resolution evolutions of the $\mathcal{A}_s=0.001$ initial data
also show significant variations in $||\,\mathcal{C}_\psi\,||(t)$ at
early times.  The time stepping error thresholds used in these runs
were the same as those used in the $\mathcal{A}_s=\{0.01,0.1\}$
evolutions which have higher overall constraint violation levels.  It
is possible that these higher resolution $\mathcal{A}_s=0.001$
evolutions could have been improved somewhat by reducing their time
stepping error thresholds.  This was not done because reducing those
tolerances would cost considerable additional computer time, and the
overall level of constraint violations in the runs presented here were
considered to be sufficiently small for the purposes of this study.

\section{Spectral Evolutions}
\label{s:Spectral Evolutions}

The spectra of the gravitational waves change in various ways as the
cosmological models evolve.  The most basic effect is the redshifting
of these waves as the spacetimes expand. The focus of this study,
however, are the more subtle effects on these spectra caused by the
non-linear interactions between the gravitational waves.  These
redshift independent effects can be extracted from the dynamical
gravitational fields by expanding those fields in harmonic basis
functions tied to the topology rather than the geometry of the spatial
hypersurfaces.  The coefficients of the harmonic basis functions in
these expansions provide spectra that are independent of the size of
the universe and are therefore independent of the redshift of those
waves.  The evolution of the gravitational wave spectra measured in
this way are caused by interactions between the gravitational waves
rather than the overall expansion of the universe.

The transverse-traceless parts of the extrinsic curvature of the $t=$
constant hypersurfaces, $K_{ij}$, are directly affected by the
gravitational wave degrees of freedom of the spacetime geometry.  The
initial data constructed in Sec.~\ref{InitialData} used this fact to
insert gravitational waves with a specified spectrum into the
dynamical fields that determine the spacetime geometry.  The evolution
these spectra are studied here by expanding the $K_{ij}(t)$ determined
by the Einstein evolution equations in a suitable basis of transverse
tensor harmonics.  The primary dynamical fields used in this study are
the fields $\psi_{ab}(t)$, $\Pi_{ab}(t)$ and $\Phi_{iab}(t)$ described
in Sec.~\ref{s:Spacetime Evolutions}.  The extrinsic curvatures,
$K_{ij}(t)$, are (somewhat complicated) algebraic functions of the
primary dynamical fields that are described in some detail in
Appendix~\ref{s:DynamicalFieldTransformations}.

The $t=$ constant spatial hypersurfaces of the spacetime models
included in this study have the topology of the three-torus.
Therefore the three-torus transverse-traceless tensor harmonics
$Y^{\,+}_{\!ij,\,\mathbf{k}}$ and $Y^{\,\times}_{\!ij,\,\mathbf{k}}$
developed in Ref.~\cite{Peng2019} are the appropriate ones to
represent the gravitational wave degrees of freedom in $K_{ij}$.
Detailed expressions for these harmonics are given in
Appendix~\ref{s:TTHarmonics}.  The tensor harmonics
$Y^{\,+}_{\!ij,\,\mathbf{k}}$ and $Y^{\,\times}_{\!ij,\,\mathbf{k}}$
form a complete basis for the symmetric transverse-traceless tensors
on the three-torus~\cite{Peng2019}.  Therefore without loss of
generality the transverse traceless parts of the time evolving
extrinsic curvature $K_{ij}(t)$ can be represented using them.  The
spectral coefficients $\mathcal{K}^{\,p}_\mathbf{k}$ that describe the
gravitational-wave degrees of freedom of the extrinsic curvature are
determined by the integrals
\begin{equation}
  \mathcal{K}^{\,p}_\mathbf{k}(t)=\int e^{ij}\,e^{mn}\,K_{im}(t)\,
  Y^{\,p\,*}_{\!jn,\,\mathbf{k}}
  \,d^{\,3}x.
  \label{e:KSpecCoefDef}
\end{equation}
These extrinsic curvature spectral coefficients are converted to
effective metric perturbation coefficients by dividing by
$2\pi\kappa$, in analogy with the initial data construction in
Eqs.~(\ref{e:PlusAmplitudeDef}) and (\ref{e:CrossAmplitudeDef}).  The
angle-averaged gravitational-wave spectrum
$\mathcal{A}{}^{\,p}_\mathrm{\,GW}(\kappa)$ is then obtained by averaging
these spectral coefficients over the tensor-harmonic modes with
parameters $\mathbf{k}$ having length
$\kappa^2=|\mathbf{k}|^2=k_x^2+k_y^2+k_z^2$:
\begin{equation}
  \left[\mathcal{A}^{\,p}_\mathrm{\,GW}(\kappa)\right]^2=\frac{
    \sum_{\kappa=|\mathbf{k}|}\,
    \mathcal{K}^{\,p}_\mathbf{k}(t)\, \mathcal{K}^{\,p\,*}_\mathbf{k}(t)}
       {4\pi^2\kappa^2\,M(\kappa)},
       \label{e:GWAmplitudesDef}
\end{equation}
where $M(\kappa)$ is the total number of gravitational-wave modes
with $|\mathbf{k}|=\kappa$.

\begin{figure}[!h]
  \centering  
  \subfigure{
    \includegraphics[height=0.3\textwidth]{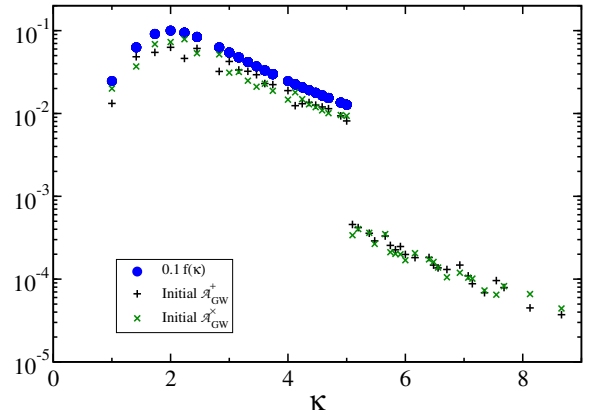}
  }
  \caption{\label{f:SpectraInitial} This figure compares the
    gravitational-wave spectrum extracted from the $\mathcal{A}_s=0.1$
    physical initial data with the spectrum of the conformal initial
    data, $0.1f(\kappa)$, used to construct that physical data.  }
\end{figure}

Figure~\ref{f:SpectraInitial} illustrates the gravitational wave
amplitudes $\mathcal{A}^{\,+}_\mathrm{\,GW}(\kappa)$ and
$\mathcal{A}^{\,\times}_\mathrm{\,GW}(\kappa)$ for the $t=0$ initial
data with amplitude scale $\mathcal{A}_s=0.1$.  For comparison this
figure also shows $\mathcal{A}_sf(\kappa)$ that defined the
gravitational wave amplitudes in the conformal extrinsic curvature.
This comparison shows that in the long wavelength domain where
$f(\kappa)>0$, the physical gravitational wave amplitudes
$\mathcal{A}^{\,p}_\mathrm{\,GW}(\kappa)$ are somewhat smaller than
$\mathcal{A}_sf(\kappa)$.  In the shorter wavelength domain where
$\kappa>5$ the physical amplitudes include an exponentially decreasing
tail that was not present in $\mathcal{A}_sf(\kappa)$.  These
differences between the initial
$\mathcal{A}^{\,p}_\mathrm{\,GW}(\kappa)$ and $\mathcal{A}_sf(\kappa)$
are caused by the non-linear process that determined the physical
initial data by solving Eqs.~(\ref{e:Kij}), (\ref{e:varphiEqn}), and
(\ref{e:TraceKConstraint}).

\begin{figure}[!h]  
  \centering  
  \subfigure{
    \includegraphics[height=0.3\textwidth]{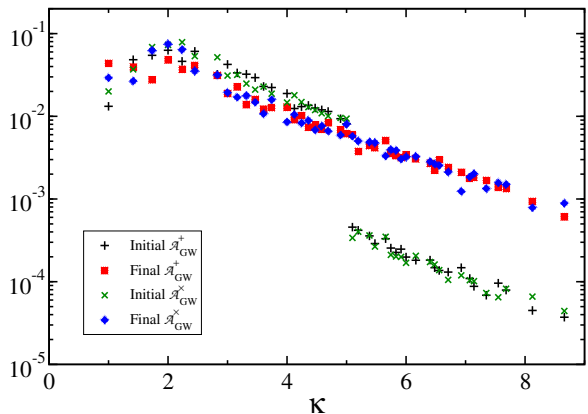}
  }
  \caption{\label{f:SpectraAmp0.1} This figure compares the initial and the
    final time gravitational-wave spectra extracted from the
    $\mathcal{A}_s=0.1$ spacetime evolutions.  }
\end{figure}

\begin{figure}[!h] 
  \centering  
  \subfigure{
    \includegraphics[height=0.3\textwidth]{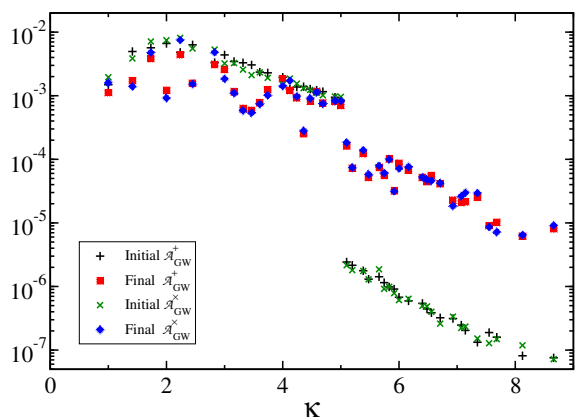}
  }
  \caption{\label{f:SpectraAmp0.01} This figure compares the initial and the
    final time gravitational-wave spectra extracted from the
    $\mathcal{A}_s=0.01$ spacetime evolutions.  }
\end{figure}

\begin{figure}[!h] 
  \centering  
  \subfigure{
    \includegraphics[height=0.3\textwidth]{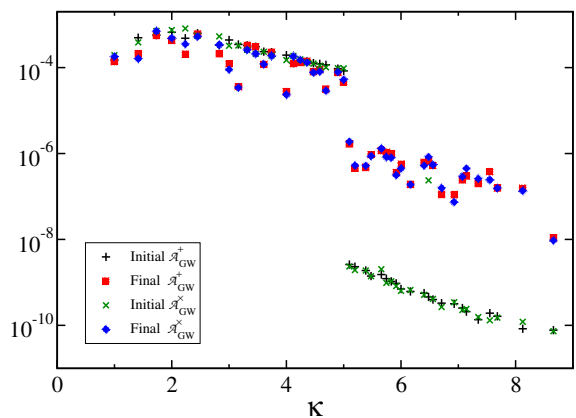}
  }
  \caption{\label{f:SpectraAmp0.001} This figure compares the initial and the
    final time gravitational-wave spectra extracted from the
    $\mathcal{A}_s=0.001$ spacetime evolutions.  }
\end{figure}

Figures~\ref{f:SpectraAmp0.1}, \ref{f:SpectraAmp0.01}, and
\ref{f:SpectraAmp0.001} illustrate the evolution of the spectra of
physical gravitational wave amplitudes,
$\mathcal{A}^{\,p}_\mathrm{\,GW}(\kappa)$, by comparing these spectra
at the initial times, $t=0$, with those spectra at the final time in
each evolution.  These results show that in almost all of the
long-wavelength range, $\kappa\leq 5$, the amplitudes at late times
show modest decreases from their initial values.  But in the shorter
wavelength range, $\kappa>5$, these results show significant
amplification of the amplitudes at later times.  This amplification is
most pronounced in the evolution of the $\mathcal{A}_s=0.1$ initial
data shown in Fig.~\ref{f:SpectraAmp0.1}.  Similar but less pronounced
amplification is also seen in Figs.~\ref{f:SpectraAmp0.01} and
\ref{f:SpectraAmp0.001} for the $\mathcal{A}_s=0.01$ and
$\mathcal{A}_s=0.001$ cases.  The evolutions included in this study
were evolved for the relatively short dynamical times needed for the
volumes of these cosmological models to increase by somewhat more than
a factor of ten.  These results show that the non-linear interactions
between gravitational waves can cause significant modifications to the
spectrum of these waves, even on the relatively short dynamical
timescales of the simulations included in this study.
%

\section{Discussion}
\label{s:Discussion}

This study has demonstrated that non-linear interactions among the
gravitational waves in expanding cosmological models can be observed
on fairly short dynamical timescales.  The cosmological models
constructed for this study were computed numerically by solving
Einstein's evolution equations using initial data containing an
ensemble of gravitational waves with different frequencies,
propagation directions and random phases.  These spacetime models were
evolved until the spatial volumes expanded by more than an order of
magnitude.  The initial data for these models were prepared with
spectra peaked at wavelengths equal to half the size of their
three-torus initial hypersurfaces.  The evolved spectra at the ends of
the evolutions in this study are shown in Figs.~\ref{f:SpectraAmp0.1},
\ref{f:SpectraAmp0.01}, and \ref{f:SpectraAmp0.001}.  These results
clearly show that energy has been transferred from the longer
wavelength gravitational waves to shorter wavelengths as a result of a
turbulent cascade in these evolutions.  These results also suggest
that the very longest wavelength modes have been amplified by an
inverse turbulent cascade.

\begin{figure}[!h] 
  \centering  
  \subfigure{
    \includegraphics[height=0.3\textwidth]{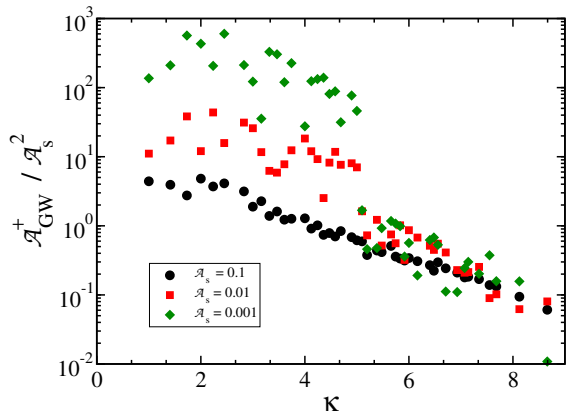}
  }
  \caption{\label{f:SpectraLateTimeComps} This figure illustrates the
    ratio $\mathcal{A}^{\,+}_\mathrm{\,GW} / \mathcal{A}_s^2$ for late
    time gravitational-wave spectra from the
    $\mathcal{A}_s=\{0.1,0.01,0.001\}$ evolutions.  This shows that
    while the late time amplitudes of the long wavelength
    gravitational waves with $\kappa\leq 5$ scale roughly linearly in
    $\mathcal{A}_s$, the short wavelength waves with $\kappa>5$ scale
    roughly quadratically in $\mathcal{A}_s$.  The analogous figure
    for the $\times$ polarization is qualitatively similar.}
\end{figure}

Figure~\ref{f:SpectraLateTimeComps} compares the late time
gravitational-wave spectra, $\mathcal{A}^{\,+}_\mathrm{\,GW}(\kappa)$,
scaled by $1/\mathcal{A}^{\,2}_s$ for the evolutions with the three
amplitude scales $\mathcal{A}_s=\{0.1,0.01,0.001\}$.  This shows that
the growth in the gravitational-wave energy density, which is
proportional to the amplitudes squared
$\left(\mathcal{A}^{\,+}_\mathrm{\,GW}\right)^2$, is roughly
proportional to $\mathcal{A}^{\,4}_s$ in the region where $\kappa>5$.
Virtually all of the gravitational-wave energy in this part of the
spectrum was generated by non-linear interactions between the waves.
Therefore the energy density in the region with $\kappa>5$ is
proportional to the dynamical-time duration of the evolutions in this
study divided by the turbulent cascade timescale $\tau_\mathrm{cas}$.
All the evolutions in this study were evolved for roughly the same
dynamical time: the time needed for the volumes of the spatial slices
to increase by about an order of magnitude.  The invariance of the
ratio
$\left(\mathcal{A}^{\,+}_\mathrm{\,GW}\right)^2/\mathcal{A}^{\,4}_s$
shown by the results in Fig.~\ref{f:SpectraLateTimeComps} therefore
implies that the turbulence cascade timescale, $\tau_\mathrm{cas}$,
observed in this study is proportional to $\mathcal{A}_s^{-4}$.  This
scaling agrees exactly with the expression for $\tau_\mathrm{cas}$
obtained in the analysis of gravitational turbulence generated by
non-linear four-wave scattering in Ref.~\cite{Galtier2017}.

The initial spectrum in Fig.~\ref{f:InitialSpectrum} was chosen to
enhance the possibility that the cascade of gravitational-wave
energies from longer to shorter wavelengths could be observed in this
study. The results displayed in
Figs.~\ref{f:SpectraAmp0.1}--\ref{f:SpectraAmp0.001} show that this
was a good choice for that purpose.  The spectrum shown in
Fig.~\ref{f:InitialSpectrum} has only a small number of
gravitational-wave modes with wavelengths shorter than its peak at
$\kappa=2$.  Therefore this initial spectrum was not an optimal choice
for exploring an inverse cascade of gravitational waves from shorter
to longer wavelengths~\cite{Green2014, Galtier2017, Ma2026}.
Nevertheless there are two results from this study that suggest the
existence of an inverse cascade.

\begin{enumerate}[leftmargin=*]

\item A careful examination of the $\kappa=1$ gravitational wave
  amplitudes for the $\mathcal{A}_s=0.1$ evolutions in
  Fig.~\ref{f:SpectraAmp0.1} shows that the late time amplitudes
  $\mathcal{A}_\mathrm{GW}^{\,+}$ and
  $\mathcal{A}_\mathrm{GW}^{\,\times}$ are larger than their initial
  values by factors of two or three.  This result is consistent with
  an expected inverse cascade.  The random phase choices used to
  construct the initial data for this study appear to be responsible
  for the gravitational-wave amplitude scatter seen in
  Figs.~\ref{f:SpectraAmp0.1}--\ref{f:SpectraAmp0.001}.  Unfortunately
  this random scatter is too large to allow the inverse cascade to be
  seen clearly in any of the other gravitational wave modes with
  $\kappa < 2$ included in this study.

\item The $\kappa=0$ spectral coefficients of the extrinsic curvature
  $\mathcal{K}_0^{\,+}$ and $\mathcal{K}_0^{\,\times}$ defined in
  Eq.~(\ref{e:KSpecCoefDef}) grow significantly during all the
  evolutions included in this study.  These $\kappa=0$
  extrinsic-curvature spectral coefficients can not be converted to
  gravitational-wave amplitudes using Eq.~(\ref{e:GWAmplitudesDef}),
  so the results can not be included in
  Figs.~\ref{f:SpectraAmp0.1}--\ref{f:SpectraAmp0.001}.  Instead the
  initial and final $\mathcal{K}_0^{\,+}$ and
  $\mathcal{K}_0^{\,\times}$ have been listed in
  Table~\ref{t:TableII}.  These results are also consistent with an
  inverse cascade of gravitational-wave energy from higher to lower
  frequencies.
\end{enumerate}
\begin{table}[!hbt]
  \caption{The average extrinsic-curvature spectral coefficients
    $\mathcal{K}^{\,+}_0$ and $\mathcal{K}^{\,\times}_0$ for the
    $\kappa=0$ modes are compared at the initial and final times for
    the evolutions of the initial data with gravitational wave
    amplitude scales $\mathcal{A}_s=\{0.1,0.01,0.001\}$.
    \label{t:TableII} }
  \begin{center}
  \begin{tabular}{lrll} 
    $\mathcal{A}_s$ & \qquad $t$ \qquad
     & \quad\qquad $\mathcal{K}^{\,+}_0$ &\quad\qquad$\mathcal{K}^{\,\times}_0$ \\ 
 \hline
 \vspace{-7pt}\\
 0.1   & \qquad $0.0$  & \qquad $2.09\times 10^{-2}$
                       & \qquad $1.89\times 10^{-2}$  \\
       & \qquad $0.1$  & \qquad $1.84\times 10^{-1}$
                       & \qquad $1.48\times 10^{-1}$  \\
 0.01   & \qquad $0.0$  & \qquad $1.09\times 10^{-4}$
                        & \qquad $8.97\times 10^{-5}$  \\
        & \qquad $1.0$  & \qquad $3.50\times 10^{-3}$
                        & \qquad $1.98\times 10^{-3}$  \\
 0.001   & \qquad $0.0$  & \qquad $1.17\times 10^{-7}$ 
                         & \qquad $9.39\times 10^{-8}$  \\
         & \qquad $10.0$ & \qquad $5.28\times 10^{-6}$
                         & \qquad $9.07\times 10^{-6}$  \\
 \hline 
  \end{tabular}
  \end{center}
\end{table}

A visual comparison of the initial and final Weyl scalars in
Figs.~\ref{f:WeylCurvature0} and \ref{f:WeylCurvature0.1} for the
$\mathcal{A}_s=0.1$ evolution suggests that the gravitational-wave
distribution becomes more spatially uniform as the spacetime
evolves. But it seems unlikely that this apparent uniformity is caused
by the growth of the lower frequency modes via the mechanism suggested
in Refs.~\cite{Green2014, Ma2026}.

The spectra at the ends of the evolutions seen in
Figs.~\ref{f:SpectraAmp0.1}--\ref{f:SpectraAmp0.001} have roughly
exponential fall off in $\kappa$ for $\kappa> 5$.  The Komolgorov
spectra for fully developed gravitational-wave turbulence would be a
power-law of the form $\sim\kappa^{-3/2}$~\cite{Galtier2017}.  While
the evolutions performed for this study could not be extended long
enough to see fully developed turbulence, there is some indication
that the exponential spectra seen in these results are flatter for the
more fully evolved larger $\mathcal{A}_s$ cases.  This might indicate
that these spectra are evolving toward power-law forms as the
turbulence becomes more fully developed.

The rapidly expanding spatial volumes $\mathcal{V}(t)$ seen in
Fig.~\ref{f:Volume} might suggest that these evolutions are undergoing
accelerated expansion or inflation. We do not believe this is the
case.  These expansions are not exponential, but instead have pole
singularities of the form $\mathcal{V}(t)=\mathcal{V}(0)/(1-t/T)^p$.
For example in the $\mathcal{A}_s=0.1$ case a numerical fit with
$T=0.142124$ and $p=2.59075$ is essentially indistinguishable from the
numerical evolution.  This type of behavior has been found to be
caused by the harmonic gauge condition used for the time coordinates
in other numerical studies of homogeneous cosmological models (see
Refs.\cite{Lindblom2026a} and \cite{Lindblom2026c}).  We believe that
this harmonic gauge condition is responsible for the pole
singularities in $\mathcal{V}(t)$ seen in the evolutions in this study
as well.

\appendix
\section{Transverse-Traceless Tensor Harmonics on the Three-Torus}
\label{s:TTHarmonics}
The three-torus can be represented using coordinates, $x^n=\{x,y,z\}$,
with ranges $0\leq x < 1$, $0\leq y < 1$, and $0\leq z < 1$, and
periodic boundary conditions.  These periodic Cartesian coordinates
are in effect the ones used in the numerical computations performed
for this study.  The transverse-traceless three-torus tensor harmonics
constructed in Ref.~\cite{Peng2019} are expressed in these periodic
Cartesian coordinates:
\begin{eqnarray}
  Y^{\,+}_{\!ij,\,\mathbf{k}}&=& \frac{e^{\,i\,2\pi\,k_nx^n}}{\sqrt{2}}
  \left(\,\hat\ell_i\,\hat\ell_j-\hat m_i\,\hat m_j\,\right),
  \label{e:PlusHarmonicDef}\\
    Y^{\,\times}_{\!ij,\,\mathbf{k}}&=&
    \frac{e^{\,i\,2\pi\,k_nx^n}}{\sqrt{2}}
    \left(\,\hat\ell_i\,\hat m_j+\hat\ell_j\,\hat m_i\,\right),
    \label{e:CrossHarmonicDef}
\end{eqnarray}
where the parameter $\mathbf{k}\neq 0$ is a vector
\begin{equation}
  \mathbf{k}=k_n=\{k_x,k_y,k_z\},
  \label{e:WaveVectorDef}
\end{equation}
whose components $\{k_x,k_y,k_z\}$ are integers.  The vectors $\hat
k_i$, $\hat\ell_i$ and $\hat m_i$ form an orthonormal triad given by
\begin{eqnarray}
  \hat k_i&=& \frac{1}{\kappa} \left\{k_x,\,\, k_y,\,\,k_z\right\},
  \label{e:khatDef}\\
  \hat \ell_i &=& \frac{1}{\lambda} \left\{-k_y-k_z,\,\,k_x-k_z,
  \,\,k_x+k_y\right\},
  \label{e:ellhatDef}\\
  \hat m_i &=& 
  \left\{\hat k_y\hat \ell_z-\hat k_z\hat \ell_y,
  \,\,\hat k_z\hat\ell_x-\hat k_x\ell_z,\,\,
  \hat k_x\hat\ell_y-\hat k_y\hat\ell_x\right\},\quad
  \label{e:mhatDef}
\end{eqnarray}
where $\kappa^2=k_x^2+k_y^2+k_z^2$ and
$\lambda^2=\ell_x^2+\ell_y^2+\ell_z^2$.

In addition to the tensor harmonics defined above that describe
propagating gravitational waves, there are five spatially independent
transverse-traceless harmonics that represent non-propagating waves.
These additional harmonics divide naturally into two classes.  The
first class consists of two time-independent harmonics with
polarizations analogous to the propagating $+$ polarization harmonics.
These harmonics are defined by
\begin{eqnarray}
  Y^{\,+a}_{\!ij,\,0}&=& \frac{1}{\sqrt{2}}
  \left(\hat y_i\, \hat y_j - \hat z_i\, \hat z_j\right),\\
  Y^{\,+b}_{\!ij,\,0}&=& \frac{1}{\sqrt{6}}
  \left(\hat y_i\, \hat y_j + \hat z_i\, \hat z_j -2 \,\hat x_i\, \hat x_j\right),  
\end{eqnarray}
where $\hat x_i$, $\hat y_i$, and $\hat z_i$ are the unit vectors in
the $x$, $y$ and $z$ directions respectively. The second class
consists of three time-independent harmonics with polarizations
analogous to the propagating $\times$ polarization harmonics.  These
harmonics are defined by
\begin{eqnarray}
  Y^{\,\times a}_{\!ij,\,0}&=& \frac{1}{\sqrt{2}}
  \left(\hat y_i\, \hat z_j + \hat z_i\, \hat y_j\right),\\
  Y^{\,\times b}_{\!ij,\,0}&=& \frac{1}{\sqrt{2}}
  \left( \hat x_i\, \hat z_j + \hat z_i\, \hat x_j \right),\\
  Y^{\,\times c}_{\!ij,\,0}&=& \frac{1}{\sqrt{2}}
  \left(\hat x_i \,\hat y_j+\hat y_i\, \hat x_j \right).
\end{eqnarray}

This collection of harmonics, $Y^{\,p}_{\!im,\,\mathbf{k}}$, provides
a complete basis for the symmetric transverse-traceless tensor fields
on the three-torus.  These harmonics satisfy the orthonormality
condition,
\begin{equation}
  \delta_{\mathbf{k},\mathbf{k'}}\,\delta_{p,p'}=
    \int e^{ij}\, e^{mn}\, Y^{\,p}_{\!im,\,\mathbf{k}}\, Y^{\,p'\,*}_{\!jn,\,\mathbf{k'}}
    \,d^{\,3}x,
\end{equation}
where the parameter $p=\{+,\times\}$ represents polarization states for
the propagating $\mathbf{k}\neq 0$ harmonics, and $p=\{+a,+b,\times
a,\times b,\times c\}$ the polarization states for the non-propagating
$\mathbf{k}=0$ harmonics.

\section{Dynamical Field Transformations}
\label{s:DynamicalFieldTransformations}
\setcounter{equation}{0} 
\renewcommand{\theequation}{B.\arabic{equation}}

The methods used in this study to solve the initial value problem, and
those used to extract the gravitational-wave spectra from the
spacetime geometry, are most conveniently expressed in terms of the
ADM dynamical fields, $g_{ij}$ and $K_{ij}$.  The evolution of the
initial data to determine the spacetime geometry, however, is most
conveniently performed using the first-order dynamical fields,
$\psi_{ab}$, $\Pi_{ab}$, and $\Phi_{iab}$.  This appendix provides a
brief summary of the relationships between these different
representations of the dynamical fields.

The transformation between the spacetime metric $\psi_{ab}$ and the
standard ADM coordinate representation of the metric is
straightforward~\cite{arnowitt1962gravitation}:
\begin{eqnarray}
  ds^2 &=& \psi_{ab}\,dx^a\,dx^b\nonumber\\
  &=& -N^2 dt^2 + g_{ij}\,\left(dx^i+N^idt\right)
  \left(dx^j+N^jdt\right),\qquad
  \label{e:ADMmetric}
\end{eqnarray}
where $g_{ij}$ is the spatial metric on the $t=$ constant
hypersurfaces, $N$ is the lapse, and is $N^i$ the shift.  The timelike
unit normal to the $t=$ constant hypersurfaces is given by
$t^a\partial_a = N^{-1}(\partial_t - N^i\partial_i)$ in these
ADM coordinates.

The transformation between the extrinsic curvature, $K_{ij}$, and the
first order dynamical fields $\Pi_{ab}$ and $\Phi_{iab}$ is more
complicated.  The standard expression for the extrinsic curvature in
ADM coordinates is 
\begin{eqnarray}
  K_{ij} &=& -\frac{1}{2N}\left(\partial_t g_{ij} -D_i N_j
  -D_j N_i\right),
  \label{e:KijDef}
\end{eqnarray}
where $D_i$ is the covariant derivative compatible with the spatial
metric, and $N_i=g_{ij}N^j$.  This expression can be re-written by
expanding the covariant derivative terms in Eq.~(\ref{e:KijDef}) into
partial derivatives and metric derivatives:
\begin{eqnarray}
  K_{ij} &=& -\frac{1}{2N}\left(\partial_t g_{ij}-\partial_i N_j
  -\partial_j N_i\right)\nonumber\\
  &&-\frac{N^m}{2N}\left(\partial_ig_{mj} +
    \partial_jg_{mi}-\partial_mg_{ij}\right).\qquad
  \label{e:KijExplicit}
\end{eqnarray}

The dynamical fields $\Pi_{ij}$ and $\Phi_{ija}t^a$ are related to the
first derivatives of the spacetime metric $\psi_{ab}$ by
\begin{eqnarray}
  \Pi_{ij} &=& - \frac{1}{N}\left(\partial_tg_{ij}-N^m\partial_mg_{ij}\right),
  \qquad\label{e:PiExpand}
  \\
  \Phi_{ija}t^a &=& \frac{1}{N}\left(\partial_iN_j-N^m\partial_i g_{jm}\right),
  \label{e:PhiExpand}
\end{eqnarray}
in a spacetime where the reference metric is the flat metric given in
Eq.~(\ref{e:ReferenceMetricDef}).  Using the expressions for $K_{ij}$
from Eq.~(\ref{e:KijExplicit}) together with the expressions for
$\Pi_{ij}$ and $\Phi_{ija}t^a$ from Eqs.~(\ref{e:PiExpand}) and
(\ref{e:PhiExpand}), it follows that $K_{ij}$ can be expressed in
terms of the first-order dynamical fields:
\begin{eqnarray}
  K_{ij} = \frac{1}{2}\left(\Pi_{ij}+\Phi_{ija}t^a + \Phi_{jia}t^a\right).
  \label{e:KijFOSHTrans}
\end{eqnarray}

The spacetime evolution code determines the fields $\psi_{ab}$,
$\Pi_{ab}$ and $\Phi_{iab}$ on a succession of $t=$ constant spacelike
hypersurfaces.  Given these fields Eq.~(\ref{e:KijFOSHTrans})
determines $K_{ij}$ in terms of $\psi_{ab}$, $\Pi_{ab}$ and
$\Phi_{iab}$ on each $t=$ constant hypersurface.  On the initial
spacelike hypersurface, $t=0$, Eq.~(\ref{e:psi0Def}) sets the lapse to
$N=1$ and the shift to $N^i=0$, so Eq.~(\ref{e:PhiExpand}) implies
that $\Phi_{ija}t^a=0$ on this initial hypersurface.  Therefore
Eq.~(\ref{e:KijFOSHTrans}) leads to the condition $\Pi^0_{ij}=2
K^0_{ij}$ given in Eq.~(\ref{e:InitialPiij}).  The condition
$\Pi^0_{ta}=0$ in Eq.~(\ref{e:InitialPita}) is a gauge choice that
sets the initial time derivatives of the lapse and shift to zero.

\section{Covariant First-Order Symmetric-Hyperbolic Einstein System}
\label{s:SymmetricHyperbolicSystem}
\setcounter{equation}{0} 
\renewcommand{\theequation}{C.\arabic{equation}}

This study uses a covariant representation of the Einstein equations
that enforces generalized harmonic gauge
conditions~\cite{Lindblom2014}.  The covariance of this representation
facilitates the study of solutions on manifolds represented using
multi-cube coordinate patches, such as the manifold with three-torus
spatial slices studied here.  All the dynamical fields in this
representation are tensors, so they can be transformed in a
straightforward way across the interfaces between coordinate patches.
The dynamical fields, $u^\alpha$, for the first-order
symmetric-hyperbolic representation of the Einstein equations used in
this study are the collection of tensor fields,
\begin{equation}
  u^\alpha = \left\{ \psi_{ab}, \Pi_{ab}, \Phi_{iab}\right\},
  \label{e:UalphaDef}
\end{equation}
were $\psi_{ab}$ is the spacetime metric, and the tensors $\Pi_{ab}$
and $\Phi_{iab}$ represent its first derivatives,
\begin{eqnarray}
  \Pi_{ab} &=& -t^c\tilde\nabla_c\psi_{ab},
  \label{e:PiDef}\\
  \Phi_{iab}&=& \tilde\nabla_i\psi_{ab}. 
  \label{e:PhiDef}
\end{eqnarray} 
The covariant derivative $\tilde\nabla_c$ used in these expressions is
the one compatible ($\tilde\nabla_c\tilde\psi_{ab}=0$) with the
reference metric $\tilde\psi_{ab}$.  This reference metric is also
used to define the Jacobians used to transform tensor fields across
the boundary interfaces between multi-cube regions.  The vector $t^c$
represents the unit timelike normal to the spacelike hypersurfaces
used to perform the evolutions.  The lower range of Latin indices,
e.g. $a$, $b$, $c$, ..., range over the four spacetime coordinates,
while the higher range of indices, e.g. $i$, $j$, $k$, ..., range over
the three spatial coordinates on each $t=$ constant hypersurface.  The
Greek indices, e.g. $\alpha$, $\beta$, $\gamma$, range over the fifty
components of the dynamical tensor fields $u^\alpha$.  The Einstein
equations for these fields can be written in the first-order form
\begin{equation}
  \partial_t u^\alpha
  + A^{k\alpha}{}_\beta({\bf u})\tilde\nabla_ku^\beta
  =F^\alpha({\bf u},{\bf \tilde\Gamma}, {\bf \partial\tilde\Gamma}),
  \label{e:EvolutionEqs}
\end{equation}
where the tensors $ A^{k\alpha}{}_\beta({\bf u})$ and $F^\alpha({\bf
  u},{\bf \tilde\Gamma}, {\bf \partial\tilde\Gamma})$ depend in
complicated ways on the dynamical fields $u^\alpha$, the reference
connection $\tilde\Gamma^c{}_{ab}$ and its derivatives
$\partial_d\tilde\Gamma^c{}_{ab}$.  The detailed expressions for this
representation are given in Ref.~\cite{Lindblom2014}.

The spacetime coordinates are determined in the generalized harmonic
representations of Einstein's equation by imposing a condition on the
difference between the connection $\tilde\Gamma^c{}_{ab}$ associated
with the reference metric $\tilde \psi_{ab}$ and the connection
$\Gamma^c{}_{ab}$ associated with the physical spacetime metric
$\psi_{ab}$.  In particular the difference between these connections
is fixed by a gauge source function $H_a$:
\begin{equation}
H_a =  -\psi_{ad}\,\psi^{bc}\left(\Gamma^d{}_{bc}-\tilde\Gamma^d{}_{bc}\right).
  \label{e:HarmonicGageDef}
\end{equation}
(Note that the difference between any two connections is a tensor.)
In general $H_a$ may be any function that depends on the physical
metric $\psi_{ab}$ (but not its derivatives) and the reference metric
$\tilde\psi_{ab}$.  The gauge condition used for the numerical
evolutions performed in this study is the simple harmonic condition,
$H_a=0$.

The generalized harmonic evolution system contains a number of
constraints.  In particular the gauge condition,
Eq.~(\ref{e:HarmonicGageDef}) is in effect a constraint
\begin{equation}
  \mathcal{C}_a =H_a +
  \psi_{ad}\psi^{bc}\left(\Gamma^d{}_{bc}-\tilde\Gamma^d{}_{bc}\right).
  \label{e:C1Def}
\end{equation}
on the dynamical fields.  In addition, the equation that defines
$\Phi_{iab}$, Eq. (\ref{e:PhiDef}), is also a constraint
\begin{equation}
  \mathcal{C}_{iab} = \tilde\nabla_i\psi_{ab}-\Phi_{iab}.
  \label{e:C3Def}
\end{equation}
These primary constraints, $\mathcal{C}_a$ and $\mathcal{C}_{iab}$,
satisfy a second-order system of evolution equations as a consequence
of the Einstein evolution system, Eq.~(\ref{e:EvolutionEqs}). (See
Ref.~\cite{Lindblom2014} for details.)  This second-order constraint
evolution system can be converted to a first-order symmetric
hyperbolic system by introducing the secondary constraints,
\begin{eqnarray}
  \mathcal{F}_a &=& t^c\nabla_c\mathcal{C}_a,
  \label{e:F1Def}\\
  \mathcal{C}_{ia} &=& \nabla_i\mathcal{C}_a,
  \label{e:C2Def}\\
  \mathcal{C}_{ijab}&=&2\tilde\nabla_{[i}\mathcal{C}_{j]ab}.
  \label{e:C4Def}
\end{eqnarray}
The symmetric-hyperbolic evolution equation for the collection of
constraints,
\begin{equation}
  \mathcal{C}^\alpha=\left\{\mathcal{C}_a,\mathcal{C}_{iab},
  \mathcal{F}_a,\mathcal{C}_{ia},\mathcal{C}_{ijab}\right\},
  \label{e:ConstraintTotal}
\end{equation}
ensures that solutions to the Einstein Eq.~(\ref{e:EvolutionEqs}) that
satisfy the constraints, $\mathcal{C}^\alpha$, on an initial surface
will satisfy them throughout the evolution of those initial data.
Numerical solutions to the equations will of course contain (hopefully
small) violations of these constraints.  This study monitors the size
of these constraint violations by evaluating a norm constructed from
the composite constraint, $\mathcal{C}_\psi$, defined by,
\begin{eqnarray}
  \mathcal{C}_\psi^2 &=& \delta^{ab}\left(\mathcal{C}_a\mathcal{C}_b
  +\mathcal{F}_a\mathcal{F}_b\right)\nonumber\\
&&  +\tilde g^{ij}\delta^{ab}\delta^{cd}\left(\mathcal{C}_{iac}\mathcal{C}_{jbd}
    +\tfrac{1}{4}\tilde g^{kl}\mathcal{C}_{ikac}\mathcal{C}_{jlbd}\right).
    \qquad
  \label{e:CpsiDef}
\end{eqnarray}
This study evaluates a norm of $\mathcal{C}_\psi$ defined by
\begin{equation}
  ||\,\mathcal{C}_\psi\,||^2 = \frac{\int \mathcal{C}_\psi^2\sqrt{g}\,d^{\,3}x}
  {\int\mathcal{N}_\psi^2\sqrt{g}\,d^{\,3}x}.
  \label{e:CpsiNormDef}
\end{equation}
The quantity that appears in the denominator, $\mathcal{N}_\psi$,
is defined by
\begin{eqnarray}
  \mathcal{N}_\psi^2&=& \delta^{ab}\delta^{cd}\tilde g^{ij}\left(
  \partial_i\psi_{ac}\partial_j\psi_{bd}
  +  \partial_i\Pi_{ac}\partial_j\Pi_{bd}\right)\nonumber\\
  &&+\delta^{ab}\delta^{cd}\tilde g^{ij}
  \tilde g^{kl}\partial_i\Phi_{kac}\partial_j\Phi_{lbd}.
  \label{e:NpsiDef}
\end{eqnarray}
It has been included in the definition of $||\,\mathcal{C}_\psi\,||$ to 
provide a dimensionless measure of the fractional errors due to
constraint violations in the numerical solutions to the Einstein
evolution system.


\acknowledgments We thank Michael Holst for providing some of the
computational resources used to perform this research.  F.Z. was
supported by the National Key Research and Development Program of
China grant 2023YFC2205801, and the National Natural Science
Foundation of China grants 12433001 and 12021003.  L.L. was supported
in part by grant No. 2407545 from the National Science Foundation to
the University of California at San Diego, USA. P.Z. was supported by
the Talent Introduction Fund grant 2020BS035 at Henan University
of Technology and the Natural Science Foundation of Henon
grant 232300421351.


\bibliography{../References/References.bib,../References/ref.bib}

\end{document}